# The relevance of learning quantum physics from the perspective of the secondary school student: A case study

Tania S. Moraga-Calderón[1], Henk Busiman[1] and Julia Cramer[1,2]
[1] Leiden Institute of Physics, Faculty of Science, Leiden University, Leiden, The Netherlands
[2] Department of Science Communication and Society, Faculty of Science, Leiden University, Leiden, The Netherlands
For correspondence: t.s.moraga.calderon.2@umail.leidenuniv.nl

**Abstract:**
Studying quantum physics in upper secondary school is now a standard practice (Stadermann *et al.*, 2019). But given the context of science education, with low recruitment numbers in higher education and poor attitudes towards science, it remains a question whether students find the learning of quantum physics relevant. In this study, we explore how students perceive the importance of quantum physics and technology and whether their perception changes after an intervention, namely the "Quantum Rules!" visit. We also aim to understand if they overall feel that learning quantum is relevant or not. In order to answer these questions, we followed a mixed-methods approach, combining both questionnaires and interviews. The quantitative analysis showed that the 'Quantum Rules!' intervention has a positive effect on students' perception of the relevance of quantum physics and technology, especially regarding how important they feel quantum science is for society. Nevertheless, the qualitative information revealed that although students may find quantum physics and technology important for society, that does not necessarily mean that they find *learning* quantum physics relevant. We found that students believe the latter is relevant to them only if they find it interesting. We therefore rediscover the common expression "important, but not for me", and we further propose that this perception derives from students not seeing the societal relevance of learning quantum physics.

**Introduction**

Quantum physics is now a standard subject in high secondary school curricula (Stadermann *et al.*, 2019). There are several reasons why countries have decided to incorporate it. Firstly, it is a fundamental and modern aspect of physics. Secondly, many applications of quantum technology are present in our modern life (lasers, microchips, solar cells) (Krijtenburg-Lewerissa *et al.*, 2017) and many more are promised for the future, such as quantum computing and quantum internet technologies (Vermaas, 2017). Thirdly, it is believed that quantum physics is a topic that fascinates students and gives physics a more attractive image (Angell *et al.*, 2004; Stadermann *et al.*, 2019).

But, do students themselves feel learning quantum physics is relevant? The decision of *what* is relevant and *why* it is so will depend to whom you ask. It has been recognized that science education does not provide a satisfactory education for the majority and 'has largely been framed by scientists who see school science as a preparation for entry into university rather than as an education for all' (Osborne and Dillon, 2008). In the case of quantum physics, the experts' view of what is relevant to teach in secondary school has already been researched (Krijtenburg-Lewerissa *et al.*, 2019). We find it necessary to consider the student voice. As has been highlighted by Rudduck and Flutter (200), we think it is important to not underestimate students' judgement and capacity of reflection, and indeed listen to what they have to say.

In this study, we explore secondary Dutch students' perceptions of the importance of quantum physics and why they should learn it, in the context of a school visit to the educational quantum lab "Quantum Rules!" based at Leiden University. Although the results were also used to develop educational material, the aim of this article is to show students' perceptions and contribute to our understanding of their point of view.

*Relevance in science education*
Science education is going through difficult times. Although young pupils are very excited when they start science in primary school, once they graduate secondary school many find science alienating (Osborne and Dillon, 2008). Student's attitudes towards science become more negative as they grow older, and students following scientific careers has decreased (Barmby *et al.*, 2008; Osborne *et al.*, 2003; Potvin and Hasni, 2014). Among others, (ir)relevance of science has been pointed out as an important factor which influences students' attitudes towards science (Barmby *et al.*, 2008; Osborne *et al.*, 2003; Raved and Assaraf, 2011).

One aspect of relevance is *personal interest*. An international quantitative study on relevance in science education was the ROSE (Relevance of Science Education) project (Schreiner and Sjøberg, 2004; Cavas *et al.*, 2009; Elster, 2007; Jenkins and Nelson, 2005; Sjøberg and Schreiner, 2010; Vázquez, 2013). The approach of ROSE was an extensive questionnaire of Likert-scale items which focused on several aspects of science relevance (interests, opinions, school science, future participations, etc.), with a special emphasis on interests. They found that interests were different between girls and boys depending on their *context* (social, theoretical, technical, ethical, etc.), as opposed to their *content* (physics, botany, chemistry, etc.). In this way, boys would generally be more interested in the technical, mechanical, spectacular, violent, explosive; whereas girls tend to be more interested in health and medicine, beauty and the human body, ethics, wonder and speculation (Sjøberg and Schreiner, 2010).

Relevance in science education can also be *societal*, although acknowledging this aspect does not necessarily motivate students to study science. The application of the ROSE questionnaire in England showed that on average students believe science "is important, but not for me" (Jenkins and Nelson, 2005). Students seem to be optimistic about the role of science in society, in that it is important and brings more benefits than disadvantages, but they do not wish to follow a future career in science. Similarly, a study about the science curriculum in England encountered the phrase: 'yes, studying science post-16 is important, but not for me' (Osborne and Collins, 2001). Pupils would acknowledge that science was behind contemporary technology and ways of living, but it was not necessarily something they wanted to learn. Furthermore, the use value of science as a societal *tool* was not recognised: 'there was little recognition that one value of scientific knowledge was the facility to engage critically with contemporary subjects' (Osborne and Collins, 2001).

Another way of understanding relevance is to see how it enables *connections*. Schollum and Osborne (1985) suggest that a science subject can be relevant if it is related to the pupils' everyday events, their existing ideas or their human relationships. Indeed, Raved and Assaraff (2011) point out in their study of secondary students' views on the learning of biology in Israel, that those students who claimed they saw no value in science were those who could not find a connection between the science subjects and their everyday life. Finding relevance of science through connections with others, or human relationships, happens when a topic becomes relevant because somebody important (family, friends, teachers) values it (Schollum and Osborne, 1985). This angle can be connected to the affective aspect of learning: 'the feeling of warmth deriving from ideas and viewpoints similar to one's friends' (Freyberg and Osborne, 1985). In fact, a study on secondary school students in Malaysia showed that the strongest factor that influenced their motivation in physics learning was relationships (Saleh, 2014). Finally, it is interesting to see how everyday life and human relationships mingle. A study on physics learning in Norway revealed that the rather common expression 'everyday life' did not actually refer to events happening concretely around the students, rather it referred to subjects or events that they would talk about (Angell *et al.*, 2004). Thus, relativity, quantum physics and astrophysics were more 'connected to everyday life' than mechanics, electricity and waves. Students (specially girls) would find science knowledge important when they could use it to engage in daily conversations, understand the world and explain it to other people (Angell *et al.*, 2004; Osborne and Collins, 2001).

A recent study reviewed the literature on relevance in science education and proposed a unified model (Stuckey *et al.*, 2013). They suggest that the concept 'relevance' in science education is multidimensional and its analysis should incorporate the idea of *consequences*, where science learning becomes relevant whenever these implications are positive for the student's life. According to this model, there are three dimensions that categorise relevance in science education (see Figure 1):

- The *individual* dimension includes learning which is relevant because it satisfies the learner's curiosity or interests, contributes to the development of their intellectual skills, and provides the student with necessary and useful skills for coping with their everyday life today and in the future.
- The *societal* dimension refers to that which is relevant because it promotes learner competency for current and future societal participation. It focuses on preparing students to be self-determined and understand the interactions between science and society, in order to become responsible citizens.
- The *vocational* dimension refers to awareness, orientation and understanding of career chances. It also encompasses any learning which opens doors for the next learning stage, in terms of coursework and achievements, for example.

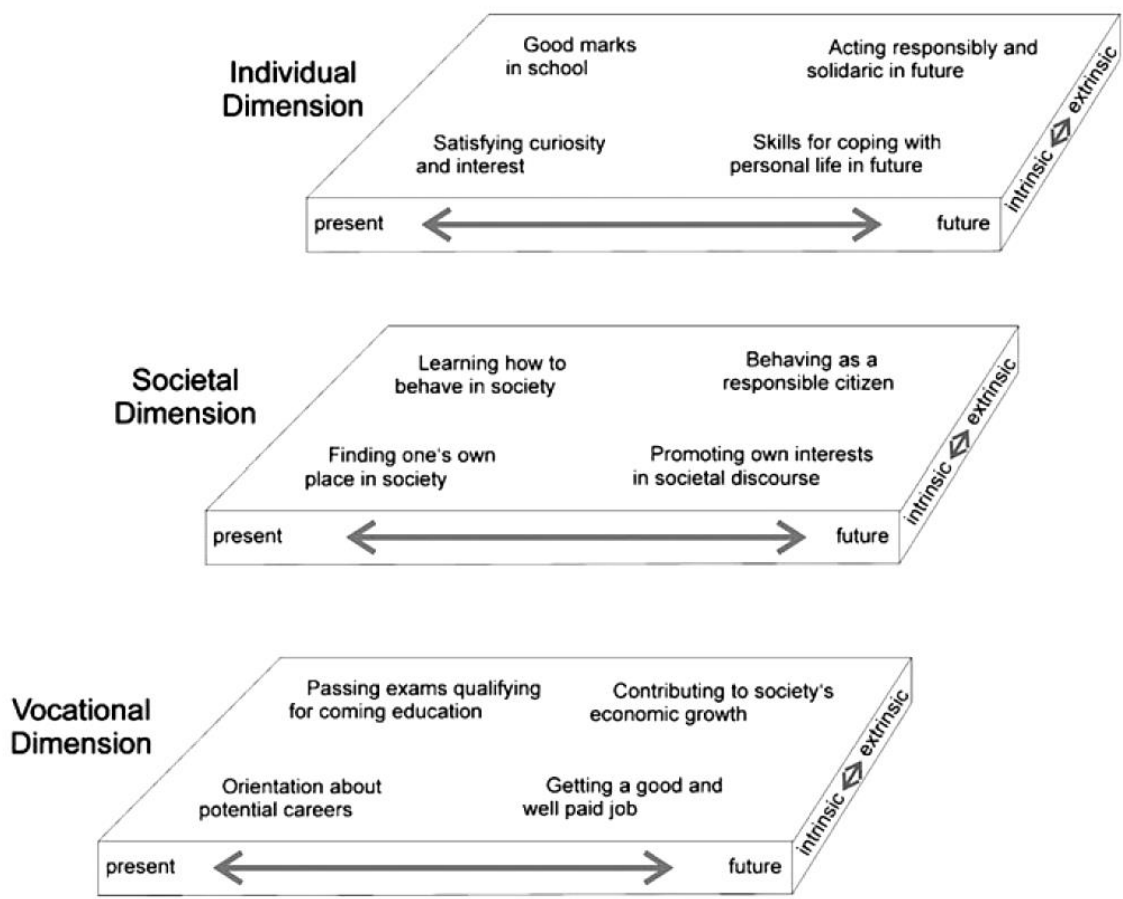

**Figure 1.** A model of relevance in science education. Image retrieved from Stuckey *et al.* (2013).

The authors point out that these dimensions are not necessarily independent and might overlap. They also explain that the relative balance between individual relevance and societal relevance varies depending on the age group of the students. Younger students might find the individual dimension more important, but as students grow and mature, this importance will shift towards societal relevance. Older students might require a science curriculum more focused on science and society than science and the individual (Newton, 1988).

*Quantum in secondary education*
Quantum physics is part of the secondary curriculum in several countries, as shown by Stadermann et al. (2019). The study included curricula from 15 countries: Australia, Austria, Belgium, Canada, Denmark, Finland, France, Germany, Italy, Netherlands, Norway, Portugal, Spain, Sweden, and the United Kingdom. In most of these countries, quantum physics is part of an elective advanced physics course for 17- to 19- year old students, which is typically taken by 5% to 20% of the overall student population (Stadermann *et al.*, 2019). Germany and the Netherlands are exceptions, where 40% to 50% of the upper secondary school students take advanced physics in their final exams (Stadermann *et al.*, 2019). In the case of the Netherlands, quantum physics was included in the curriculum ten years ago to their highest secondary educational variant VWO (Voorbereidend Wetenschappelijk Onderwijs). Students usually cover this unit during their last year of secondary school (6VWO), when they are 17-18 years old. The aim of the reform was 'to promote scientific skills and thinking, and provide a vision on the relevance of science and technology in society' (Commissie Vernieuwing Natuurkundeonderwijs, 2010).

The student voice in quantum education research can be found in studies about identity and science culture (Bøe *et al.*, 2018; Johansson *et al.*, 2018). A recent study of the ReleQuant project in Norway showed that secondary school students who had used their material have very much in common with what they call the "implied student of the traditional physics classroom": identifies him/herself as a physics student by performing well in tests and final qualifications, sees learning physics as the main goal of physics education and is motivated by interest in the subject itself. In contrast, students did not see history and philosophy of science aspects as learning goals in themselves and would feel uncomfortable in situations like group discussions, where they would struggle to determine how well they had performed (Bøe *et al.*, 2018). Similarly, Johansson et al. (2018) recognized that the ways of being "a good physics student" were limited by the dominating focus on calculating quantum physics. They argue that this narrow discursive position limits the possible available ways of being a physicist and is likely to exclude students who do not feel identified with it.

These last studies do talk about relevance, in a way. Some of their discussions refer implicitly to the questions: "is quantum physics important?" and "why do I learn quantum physics?". The main reason these students find quantum physics relevant is because they are very intrinsically motivated by it: they find it fascinating and interesting (Bøe *et al.*, 2018; Johansson *et al.*, 2018), and they will pursue learning it even when it is challenging, because it defines them as a good student to achieve well in a difficult physics subject such as quantum physics. Furthermore, the traditional physics classroom reproduces this type of student (Johansson *et al.*, 2018).

*The "Quantum Rules" visit: quantum outreach as a complement to school education*
Outreach activities in the context of secondary school are a mixture between informal and formal education (Jarman, 2005; Vennix *et al.*, 2018). On one hand, outreach activities may be considered informal, as they occur outside school and there is often freedom to choose whether to attend or what to work on (Vennix *et al.*, 2018). On the other hand, learning within these activities is still structured and has a desired outcome (Jarman, 2005). These short interventions, if designed well, might enhance the learning of specific aspects in the school curriculum and enhance students' positive attitudes towards science (Potvin & Hasni, 2014; Choudhary *et al.*, 2018).

The setting for our research is the "Quantum Rules!" lab, a project permanently based at the Faculty of Science, Leiden University. The lab contains around 20 quantum experiments, all designed according to the Dutch quantum curriculum. Each year, around 30 school groups from the last year of upper secondary (6VWO, 17-18-year-old) school students come for a day visit.

All "Quantum Rules!" visits follow a similar structure, which varies slightly in which experiments are being done or how long the different activities take, depending on the size of the group, the school, or any other specific preference of the school teacher. The structure of a typical visit can be seen in Table 1. The presentations are usually coordinated so that content-related experiments present right after the other or even in one same presentation, allowing connections between the physics topics to appear.

Table 1. The "Quantum Rules!" intervention

| Time | Activity | Description |
| --- | --- | --- |
| 2 hr 30 min | Experiments | The students choose an experiment and work in pairs autonomously. They are supported by teachers and university students. |
| 45 min | Lunch Talk | The students attend a 20 minutes lecture given by a physics or astronomy student. |
| 1 hr | Presentations | The students explain to their classmates how they did the experiments, their main findings and reflections. |

During the "Quantum Rules!" intervention students work autonomously and choose what experiment to perform, there is a strong practical work component, students do the activities with their classmates and there is contact with experts. Moreover, the visit is focused on *understanding* quantum physics, as its main objective is to prepare the students for their final exams. These properties enhance autonomous motivation and are likely to generate positive attitudes in students (Vennix *et al.*, 2018). Therefore, we think that the "Quantum Rules!" visit is a suitable environment to explore how and why attitudes change, and in particular, to explore the changes of the "Importance of (quantum) science" construct (Barmby *et al.*, 2008).

*Aim and research questions.*
Considering what is known about relevance in science education, the fairly recent incorporation of quantum physics in the secondary school curriculum (6VWO students) and the "Quantum Rules!" experience as an example of a complementary activity to school education, our study aimed to reveal students' point of view about the relevance (or importance) of quantum science, how it may change after a short intervention and why.

We distinguish the *importance* of quantum science from the *relevance of learning* quantum physics. The former answers the question "Is quantum science important?" and might be answered from any viewpoint (for me, for society, etc), whereas the latter answers the question "Why should *I* learn quantum physics?" and will be necessarily answered from a personal standpoint. We wish to address both aspects in this study.

Concretely, we formulated the following research questions:
- Does students' perception of the importance of quantum physics and technology change after a "Quantum Rules!" visit?
- What dimensions of relevance, according to the model of Stuckey *et al.* (2013), do students recognize in their learning of quantum physics?
- Do these 6VWO students feel learning quantum physics is relevant?

**Methodology**

An overview of the methodology can be seen in Table 2. In order to answer the research questions, we adopted a mixed-methods approach, with both quantitative and qualitative components. This mixed methodology enables us to build upon the research of previous studies while still exploring new perspectives.

The data was collected in two stages: i. the application of self-reported questionnaires before and after the visit (pre- and post-tests) of 45 students from the same school distributed in three different visits, and ii. interviews to four of those students a month later.

**Table 2.** Methodology overview.

| Method | Instrument | Description |
|---|---|---|
| Quantitative | Questionnaire, section A: *'What I want to learn about'* | 12 Likert-type items with 4 categories, related to students' **interests** in topics related to quantum physics. |
| | Questionnaire, section B: *'My opinions about quantum science'* | 9 Likert-type items with 4 categories, related to students' **opinions** on how quantum science has an impact on society. |
| Qualitative | Questionnaire, section C: *'Why do I learn quantum physics'* | An open question aimed at understanding **why** students learn quantum physics, with the format *I think learning about quantum physics …. because ….* |
| | Interviews | 20-minute interviews in pairs, focused on **deepening** the findings of the questionnaire application. |

*Questionnaires*

Our questionnaire was heavily inspired by the ROSE questionnaire (Schreiner and Sjøberg, 2004). One of the strengths of this questionnaire was that it was designed in a playful and concrete way, such that it would connect to the pupils' view. The main idea behind our developed questionnaire is that the importance of quantum has at least an individual component (interests) and a societal component (opinions). To connect to other dimensions of relevance (see Figure 1), we decided to include an open question as well. Consequently, the structure of our survey consists of three sections:

A. *What I want to learn about*. This section contains 12 Likert-type scale items, each with four categories from 'Not interested' to 'Very interested' (the two middle categories did not have labels). The neutral 'middle' category was left out on purpose, since it does not have the same meaning for all people (Schreiner & Sjøberg, 2004). This section is a quantum-adapted version of the 'What I want to learn about' section in ROSE. Some of the statements are related to specific experiments in the lab, such as A1 'How does light generate electricity' (related to the photoelectric effect experiments) and A3 'How X-rays, ultrasound, etc. are used in medicine' (related to the PET scan experiment). Statements A2, A3, A4 and A6 were taken directly from the ROSE questionnaire (Schreiner & Sjøberg, 2004).
B. *My opinions about quantum science.* This section contains 9 Likert-type scale items, again each with four categories from 'Disagree' to 'Agree' (the two middle categories did not have labels). On addition to the reason mentioned above, we decided to not include a 'neutral' category because we thought it could tend to be a comfortable option for the students. Most of the statements from this section were adapted from the ROSE section 'My opinions about science and technology'.
C. *Why do I learn quantum physics.* This section was an open question were students had to complete the sentence '*I think learning about quantum physics* (blank) *because* (blank)'. The second part of the sentence could be as extensive as the student wanted. As already stated above, the objective of this question was to open new doors for other kinds of reasons for the (lack of) importance of quantum and the learning of it.

Once the design was complete, the questionnaire was first translated from English into Dutch using the online tool Google Translate. Thereafter, the translated version was revised and corrected by three native Dutch-speaking people. A pilot study was performed successfully with a group of nine Dutch students. The final version can be found in Appendix A.

The questionnaire was applied to three consecutive visits of around 15 to 20 students each, which made a total of 45 students. All the students who came to the visits agreed to participate. The visits were from the same school, but each had a different physics school teacher. The pre-test was answered at the beginning of the day, whereas the post-test was applied after the afternoon presentations were over (see Table 1).

*Interviews*

We performed two interviews with two students each. The students who participated were all volunteers. The interviews were semi-structured and contemplated both direct questions and controversial statements (see Appendix B). In order to avoid immediate consensus in the case of statements, the students were asked to first write their answers on a piece of paper and later share them with the group, as suggested by Osborne and Collins (2001). Interviewing two students at the same time allowed certain degree of discussion and gave the atmosphere of a conversation. It was also useful in terms of language, as they were done in English and sometimes one of the (Dutch) students would ask the other how to translate a specific concept. Each interview lasted around 20 minutes and was performed at the students' school approximately a month after their visit, due to practical constraints. The interviews were recorded and later transcribed by the main author, where all names were changed into fictional ones to ensure anonymity.

*Data analysis*

Our study included quantitative data from sections A and B and qualitative information from section C and the interviews. Here we detail how we analysed each of these components.

*Quantitative analysis*. The data from sections A and B was analysed using SPSS (Statistical Package for the Social Sciences). To assess differences between the pre- and post-tests, we used the paired-samples sign test. This statistical test was selected because of the non-parametric ordinal character of the data, and because the distribution of the difference was neither normal nor symmetric. The null hypothesis (there is no difference between pre- and post-tests) was rejected when the reported p-value was less than 0.05. We also performed Mann-Whitney U-tests to assess the differences between girls (n=24) and boys (n=21), although given the reduced sample size, we considered significant differences when the p-value was smaller than 0.01.

*Qualitative analysis.* The main qualitative data were the answers from the open question of the questionnaire (section C). Although the interviews proved to be very interesting, due to the low number we did not reach saturation of the information. Therefore, we used quotes from the interviews to deepen and prioritize patterns which were already present in the answers of the whole group of students.

The qualitative data from section C was coded according to two categories: *value* and *dimension*. The category value was coded considering only the first part of the sentence ('I think learning about quantum physics ...'), whereas the category dimension was coded considering the entire statement. The category *value* was designed in an inductive way, by observing the patterns in the data and clustering according to what we thought could help us allow structure it better. On the other hand, the category *dimension* was defined in a deductive way, according to the three dimensions of relevance in science education of Stuckey *et al.* (2013). The codebook is described in Table 3.

There are two comments regarding our codebook. Firstly, at the beginning statements of the sort '*I think learning about quantum physics is* difficult / useful but complicated / complex …' were coded as 'Negative' in the value category. But upon further analysis, we realised that this was a prejudiced idea, and that finding a subject difficult does not necessarily mean the person does not find the subject important. Therefore, those kinds of statements were coded as 'Missing'. Secondly, although we initally coded the 'Societal' dimension following the definition of Stuckey *et al.* (2013), we finally also included societal expressions that did not directly refer to societal *participation*. We will further explain this in the following section.

Table 3. Codebook for the open question in section C of the questionnaire.

| Category | Code | Explanation |
| --- | --- | --- |
| **Value** *'I think learning quantum physics …'* | Positive | Is interesting, useful, important |
| | Neutral | Can be useful, interesting for some |
| | Negative | Is useless, not interesting, not important |
| | Missing | Does not exist or is not clear. Also includes 'difficult'. |
| **Dimension** *'I think learning quantum physics … because … '* | Individual | Personal intellectual skill development or satisfaction of curiosity. Related to: interest, understanding, usefulness in daily life, consequences for personal life. Ex: *'… is not very useful for us, because it is very deep and we are not going to use it in our daily lives / it does not change our way of thinking.'* [Pre, ID 113] |
| | Individual and societal | Both individual and societal components. For example, both interesting and useful for society. Ex: *'… can be useful because it can make our society better, but I think you should only learn it if you are very interested in it.'* [Post, ID 101] |
| | Societal | Learn competency for current and future societal participation. The statement connects the learning of quantum to societal issues or applications. Ex.: *'… can be useful because it can bring society forward, but it is also important to know its dangers.'* [Post, ID 102] |
| | Vocational | Vocational awareness and understanding of career choices. Includes: learning enough to be able to choose, pass exams to go to further education. Ex: *'… is important to train students as broadly as possible so students can experience where their interest lies.'* [Pre, ID 112] |
| | Missing | There is not enough information to code a dimension. Ex. *'… is complicated but useful, because it comes back everywhere.'* [Post, ID 220] |

The coding was performed by all three authors. Since the original answers were in Dutch, any misunderstanding due to language was discussed with all authors. The coding process was done in three rounds. On the first one, there was a high level of disagreement, especially regarding the category dimension. A second round was performed where each coder revised his/her decisions, after which there was 66% agreement. Full consensus was finally reached in a face-to-face meeting where all disagreements were discussed and resolved.

**Results and Discussion**

In this section we show our main results after the complete analysis. We have divided this section into two subsections:
1. *The effects of the intervention.* Here we reflect upon the differences between pre- and post- tests. We find that students' interests are stable, whereas students' opinions change more. Students significantly agree more with the statement *"Quantum science is important for society"* after the visit and see the learning of quantum physics as more valuable.
2. *'Important, but not for me'*. Students believe the most important dimension in learning quantum physics is the individual one: someone *interested* in it will be willing to understand it. Consequently, if students are *not* interested in quantum, they do not see the relevance of learning it. For them, the societal relevance of learning quantum physics is not strong enough to constitute a reason for learning.

*The effects of the intervention*
The main results of section A 'Interests' and section B 'Opinions' can be seen in Figure 2. Both interests and opinions are more positive or optimistic after the visit, at least for the statements which change significantly. Furthermore, opinions appear to vary more than interests. Indeed, on average 3 of 12 (25%) interest statements were significantly perceived as more interesting after the visit, whereas 5 of 9 (56%) opinion statements were significantly perceived as more agreeable.

Of the three interest statements which changed significantly, two are related to experiments in the "Quantum Rules!" lab. Statement A1 'How light can generate electricity' is related to the photoelectric effect experiments, whereas statement A3 'How X-rays, ultrasound, etc are used in medicine' can be linked to the PET-scan experiment. It is peculiar that statements A11 and A12 are so unpopular. It could be that the students coming to the "Quantum Rules!" visit are socialized within the "traditional physics classroom", and therefore do not see nature of science or history and philosophy of science as learning goals in themselves (Bøe *et al.*, 2018).

Regarding opinions, students on average appear to have both optimistic and pessimistics views. On average, they seem to relate quantum technologies to easier and more comfortable lives (B2) and environmental problems (B6), but don't see how it may help eradicate poverty (B5). Not only is the mean average quite low for both pre- and post-tests, but three students marked it as a missing value, indicating with question marks or a few words that they couldn't make the connection. Regardless, of all statements, B1 "Quantum science is important for society" was the one whose null hypothesis was rejected the strongest, with a p-value smaller than 0.001. This strongly suggests that, on average, students think science is more important for society after the visit.

Some differences arise when comparing results between girls and boys (see Appendix C). Boys were more interested in explosive events, such as the atomic bomb (A6), whereas girls were more interested in health-related issues, like the mutations of DNA (A9). Furthermore, girls were more critical, or sceptical, of whether the statements in section B and would choose not to answer them, especially in the pre-test. These two differences had been found in previous ROSE studies (Jenkins, 2006; Sjøberg & Schreiner, 2010).

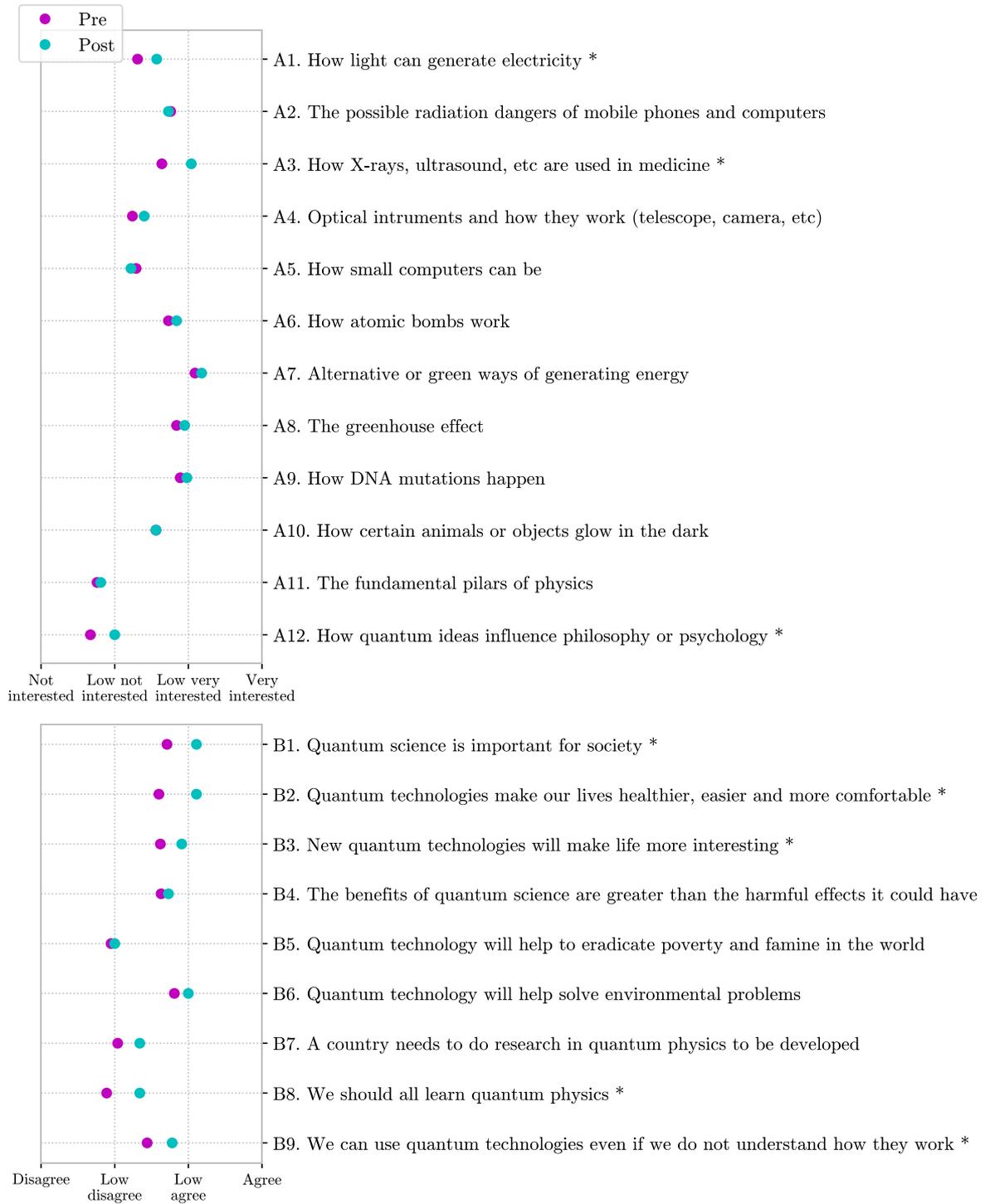

**Figure 2.** Mean of interests (A-statements) and opinions (B-statements) in pre- and post-tests. The labels "Not interested"/"Disagree" and "Very interested"/"Agree" were indicated in the questionnaire, whereas the middle labels were not. In order to calculate means and perform statistical tests, the options where transformed into numerical values of "1, 2, 3, 4" in ascending order of interest/agreement. Statements A1 (p-value 0.007), A3 (p-value 0.004), A12 (p-value 0.031), B1 (p-value lower than 0.001), B2 (p-value 0.035), B3 (p-value 0.043), B8 (p-value 0.017) and B9 (p-value 0.007) were statistically different in pre- and post-tests when performing paired-samples sign tests.

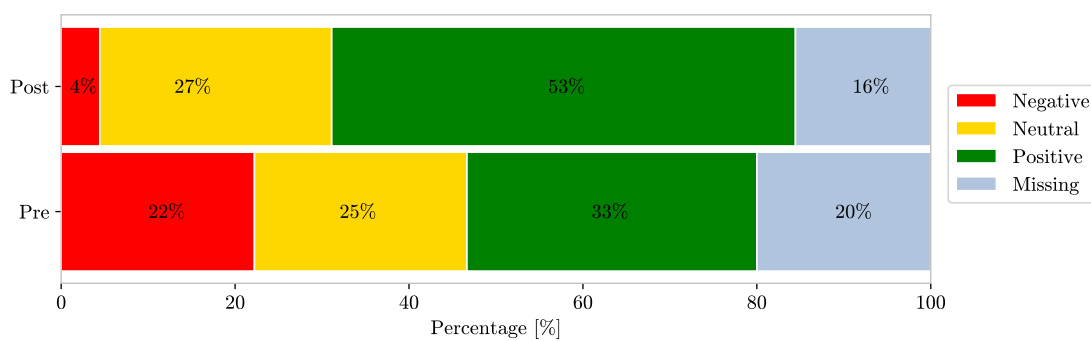

**Figure 3.** 'Value' coding of the answers to section C '*Why do I learn quantum physics?*' in the pre-test (below) and post-test (top). Only the first part of the statements was used to code this category (*I think learning about quantum physics…*). The codebook is explained in Table 3.

Apart from considering quantum science more important for society, on average students find the learning of quantum more meaningful. The coding of the category "Value" (see Table 3) of section C 'Why do I learn quantum physics?' shows that students describe 'learning quantum' as more useful, interesting or important after the visit (Figure 3).

Regarding how students would individually change their perception of value, it is interesting to note that no student expressed that the learning of quantum after the visit was *more* negative. No student went from a 'difficult' perspective (coded as "Missing") to a negative one. Of the students that negatively viewed the learning of quantum at the beginning of the visit, only two still considered quantum physics useless or unimportant after the visit. These students would refer to quantum physics' lack of usefulness or lack of connection to daily life. In this sense, the lack of connection to everyday life seems to be the argument students use when they find no relevance in a scientific topic (Raved and Assaraf, 2011).

The concept of everyday life thus proves to be highly important, despite (or because of?) the ambiguity of it. As other studies have pointed out, it is not clear what daily life really means, as students may actually refer to their daily conversations more than to the daily objects or phenomena that surround them (Angell *et al.*, 2004). We think that human relationships play a role in defining that which is everyday life, as mediators perhaps, of that which is strange to that which is familiar. It could even be that the connection to daily life enables a more abstract connection to the whole of society. A student in the interviews explained how hearing a woman talk about her experience allowed him to see what quantum does for the world:

> B: Definitely the fact that we also learned a bit about what quantum does for a country or for the world changed my view of it.
> *I: Which element of the visit gave you that connection to the rest of the world?*
> B: Our instructors, they just told us their experiences… we had a presentation of a woman who did research on a telescope and… yeah, that probably helps a lot, like hearing others talk about quantum and what they have done with it in their lives or why they are so interested in it.
> <div align="right">*Interview 1*</div>

*'Important, but not for me'*
Although students on average do find quantum science important for society, this does not mean that they think it is relevant that *they* learn it. Indeed, the expression 'Important, but not for me' is again found in our study, as has been reported previously (Jenkins and Nelson, 2005; Osborne and Collins, 2001). The results from the coding of the 'Dimension' category helped us to better understand this phenomenon (see Figure 4).

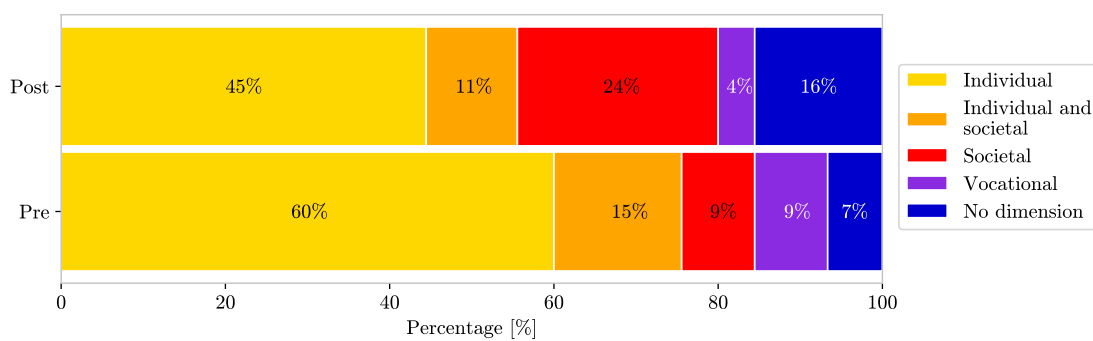

**Figure 4.** 'Dimension' coding of the answers to section C '*Why do I learn quantum physics?*' in the pre-test (below) and post-test (top). The whole statement '*I think learning about quantum physics... because...*' was considered for this category. The codebook is explained in Table 3.

*Students think "it is relevant to learn quantum if you're interested in it".* One of the first impressions from Figure 4 is how predominant the 'individual' dimension is, in both pre and post-tests. If one considers both 'individual' and 'individual and societal' statements, 76% of the total are related to the individual dimension in the pre-test, and 56% in the post-test. The reason behind this seems to be that, according to the students' point of view, quantum physics is so difficult that only wanting to satisfy your own personal curiosity is sufficient motivation:

> A: I think it is very important that we learn about these (quantum) things, but I think that you have to be interested in order to be willing to understand it. I don't think quantum is for everyone, a lot of people won't find it interesting.
>
> *Interview 1*

*Students do not see that quantum literacy is a useful tool for their future as citizens.* Although we did classify certain statements as 'Societal', we believe the societal dimension of learning quantum physics is not strong. Most of the statements just made explicit the connection between quantum physics and society. The students would not explain why it was relevant for *them* to learn quantum, rather they explained why quantum is important for society. For example, a student answered in the open question:

> *I think learning about quantum physics* is important *because* it can produce new technologies that can advance the lives of people and society.
>
> *Pre, ID 210*

A more sincere argument for societal relevance would be to understand the implications that scientific literacy will have for their voice as future citizens. Namely, to see how learning (quantum) science would help them get involved and have a say in current and future controversial topics. Students can see this kind of relevance in other fields. For example, in an interview a student explained how learning Dutch language might not be interesting for him, but is important nevertheless so that he knows how to behave in society:

> D: People can use language in all kinds of ways, also to influence people... and I think that in order to make educated decisions about things you need to be able to have a good understanding of the language and different techniques that people use (…) I think it is useful to know this in order to be critical. But it's not very interesting (laughs).
>
> *Interview 2*

No statement from section C of the questionnaire 'Why do I learn quantum physics?' nor quotes from the interviews spoke of this aspect of relevance contextualised in the learning of quantum science. In other words, none of these students expressed how quantum-scientific literacy would allow them to develop their own voice as citizens and judge or participate in future debates.

*Acknowledging the technological applications of quantum is not enough to see the societal relevance of learning it.* Finally, we would like to give an explanation for the presence of more 'Societal' statements in the post-test. Students were able to make more connections to society by seeing, and in occasions even experimenting with, technological quantum applications. The "Quantum Rules!" lab has experiments that relate to solar panels, the greenhouse effect and PET scan. Nevertheless, for the more critical student technological applications are not enough. This may be because these statements do not really answer why do *I* have to learn quantum, rather they answer why it is important that *somebody* learns quantum science:

> *I think learning about quantum physics* is not very important *because* it is only useful if you really have to make products or machines with it.
>
> Post, ID 114

**Summary and future prospects**

The present study aimed to understand students' perceptions of the importance of quantum science and technology, their views on whether it is relevant for them to learn quantum physics, and whether any of these changed after an intervention such as "Quantum Rules!". We used a mixed-methods approach, applying self-reported pre-post questionnaires to three visits (n = 45) and interviewing four students a month later.

The main limitation of this study is its scale. Our results cannot directly be generalised to describe the perceptions of secondary school students in the region of Zuid Holland, least to say The Netherlands. We could also see how the implementation of the questionnaire at the beginning of the day predisposed the students to think about quantum science in society, and thus possibly made them reflect more on these issues than a "normal" visit. In this respect, we face a "measurement" problem similar to that of quantum physics itself.

Although our study is just a case study, limited by its small scope, through a questionnaire based on the ROSE survey and using the model of relevance suggested by Stuckey *et al.* (2013), we have reached conclusions which are confirmed in earlier studies of science education and implied in studies of secondary quantum education.

Students do find quantum science important for society, especially after the visit (see Figures 3 and 4). This result was expected. Complementary outreach interventions such as "Quantum Rules", where the focus is on understanding, there is autonomous work with classmates, contact with experts, etc (Vennix *et al.*, 2018) are expected to improve students' attitudes towards science, including their perception of how important science is.

Nevertheless, we re-discover the expression "important, but not for me" (Jenkins and Nelson, 2005; Osborne and Collins, 2001). Students think the main reasons that justify the learning of quantum physics are those related to the individual dimension of relevance: either that quantum is interesting in itself or that it allows to understand physics better. In other words, students argue: "if I'm not interested, I shouldn't learn it".

In this sense, our result aligns well with previous findings. The students who come to the "Quantum Rules!" experience are socialized within the standard educational system and see that the teaching of (quantum) physics is targeted towards a student that wants to learn physics and is motivated by interest in the subject itself (Bøe *et al.*, 2018; Johansson *et al.*, 2018).

Perhaps more concerning is the implied result that students appear not to see the societal relevance in learning quantum physics at all. Although we did code statements as "Societal" (see Figure 5), closer inspection of those statements showed that not one of them mentioned that learning quantum is relevant because it allows to engage critically with contemporary scientific issues. Most statements, instead of giving reasons for why they should learn quantum physics, would state reasons why quantum physics is important for society in general.

In this regard, acknowledging the (technological) applications that quantum physics may have in society is important, but it is not enough. We must give students tools to see the connections between (quantum) science and their everyday life (Raved and Assaraf, 2011), see how it will allow them to make better decisions as citizens, how it can provide them models or concepts that may serve as analogies to face other problems of life (Colletti, 2019), how it may help them bond and make connections with others (Freyberg & Osborne, 1985).

It would be interesting, in a further study, to explore better what "everyday life" really means (Angell *et al.*, 2004). How is the relevance of human relationships, of conversations and respect for others, related to everyday life? Furthermore, how does this kind of relevance fit into the model of Stuckey *et al.* (2013)? We suggest human relationships and (a component of) everyday life may perhaps be part of a 'middle' layer, somewhere in between the individual and society.

The presented results may not come as a surprise and already multiple efforts have been made to incorporate the 'science, technology and society' view into science education. An important drawback to these initiatives seems to be assessment: curricula with 'nature of science' or 'history and philosophy of science' learning objectives are very hard to assess in a standardized manner (Stadermann, 2018) and students do not see them as learning goals in themselves (Bøe *et al.*, 2018). But an intervention such as "Quantum Rules!", in the boundary between formal and informal education, offers a free setting which does not have, by definition, any fixed assessment and students come with the predisposition of living a relaxed, open-ended experience. Could this environment be more suitable to explore the societal relevance of (quantum) science?

"Quantum Rules!" is currently living a reform in which a new element will be added to the experience: a collective game which will help connect better the experiments with applications and controversies in society. It would be interesting to see whether this element affects students' perception of relevance of quantum physics and contrasts the findings with the results of this article.

**Acknowledgements**

This work is part of the NWA research programme Quantum/Nano Revolution with project number 400.17.607, which is financed by the Dutch Research Council (NWO). T. S. Moraga Calderón gratefully acknowledges the financial support of CONICYT through Becas Chile 2017, Contract number 73181588. The authors would like to thank Pedro Russo for his valuable comments.

**Appendix A: Questionnaire in English**

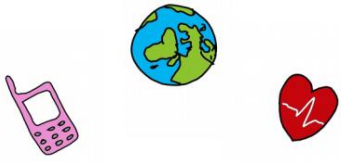 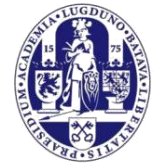

# RELEVANCE OF QUANTUM SCIENCE

This questionnaire asks you about your interests in quantum physics and technology and your perception of its importance in society.

*There are no correct or incorrect answers, only answers that are right for you.*

The information you give will be used in a small study which will help us improve the "Quantum Rules!" experience and better understand how young people perceive the learning of quantum physics.

Your answers are anonymous, so please do not write your name on this questionnaire.

**THANK YOU!**
Your answers are essential for this study.

---

**START HERE:**

I agree to participate. *(please tick the box)* ☐

My personal code is ☐☐☐

I am
☐ female
☐ male
☐ prefer not to answer

## A. What I want to learn about

**How interested are you in learning about the following?**
(Give your answer with a tick on each line. If you do not understand, leave the line blank.)

*Not interested    Very interested*

1. How light can generate electricity ……………………………………………….. ☐ ☐ ☐ ☐
2. The possible radiation dangers of mobile phones and computers ……………… ☐ ☐ ☐ ☐
3. How X-rays, ultrasound, etc. are used in medicine .................................. ☐ ☐ ☐ ☐
4. Optical instruments and how they work (telescope, camera, etc.) ……………. ☐ ☐ ☐ ☐
5. How small computers can be …………………………………………………. ☐ ☐ ☐ ☐
6. How atomic bombs work ……………………………………………………. ☐ ☐ ☐ ☐
7. Alternative or green ways of generating energy ……………………………… ☐ ☐ ☐ ☐
8. The greenhouse effect ………………………………………………………. ☐ ☐ ☐ ☐
9. How DNA mutations happen ………………………………………………… ☐ ☐ ☐ ☐
10. How certain animals or objects glow in the dark …………………………… ☐ ☐ ☐ ☐
11. The fundamental pillars of physics ………………………………………… ☐ ☐ ☐ ☐
12. How quantum ideas influence philosophy or psychology ……………………. ☐ ☐ ☐ ☐

## B. My opinions about quantum science

**To what extent do you agree with the following statements?**
(Give your answer with a tick on each line. If you do not understand, leave the line blank.)

*Disagree    Agree*

1. Quantum science is important for society ……………………………………….. ☐ ☐ ☐ ☐
2. Quantum technologies make our lives healthier, easier and more comfortable ... ☐ ☐ ☐ ☐
3. New quantum technologies will make life more interesting .................................... ☐ ☐ ☐ ☐
4. The benefits of quantum science are greater than the harmful effects it could have .......................................................................................................... ☐ ☐ ☐ ☐
5. Quantum technology will help to eradicate poverty and famine in the world …… ☐ ☐ ☐ ☐
6. Quantum technology will help solve environmental problems ………………… ☐ ☐ ☐ ☐
7. A country needs to do research in quantum physics to be developed ……………. ☐ ☐ ☐ ☐
8. We should all learn quantum physics …………………………………………… ☐ ☐ ☐ ☐
9. We can use quantum technologies even if we do not understand how they work ☐ ☐ ☐ ☐

## C. Why do I learn quantum physics

Let us know in a few sentences what you feel about learning quantum physics and why.

I think learning about quantum physics is ……………………………………………………………………………………………………..

Because ……………………………………………………………………………………………………………………………………………

…………………………………………………………………………………………………………………………………………………………

…………………………………………………………………………………………………………………………………………………………

………………………………………………………………………………………………………………………………………………………...

**Appendix B: Interview Guidelines**

Introduction   *(2 min)*
                Present myself and the project.
                State objective of the focus group.
                Ask for consent.

Discussion   *(16 min)*
                *The value of quantum science and technology (2 min)* \*
                Q1     Why, if at all, should quantum physics be studied? Why does it matter?

                *Why do I learn quantum physics (3 min per statement)* \*\*
                S1     "Quantum physics is fascinating."
                S2     "I think you should learn quantum physics only if you are very interested in it."
                S3     "Quantum science is crucial for the development of our country."

                *The "Quantum Rules!" visit (5 min)* \*\*\*
                Q2     Did your perception of relevance change after the "Quantum Rules!" visit? What elements contributed to enhance or diminish the importance of quantum? Give examples.

Closure   *(2 min)*
                Ask the participants to reflect on what is the most important issue that emerged.
                Thank for the participation.

\**Ideas to deepen the discussion*
Something may (not) be relevant because:
    - it has (not) instrumental value, it is (not) useful
    - it has (not) a beauty value or does (not) satisfy curiosity
    - it does (not) allow you to understand yourself better
    … and others.

\*\* *Other possible statements:*
"Learning quantum arouses interest in areas that are important for the development of our country."
"Quantum is the future."
"I find quantum physics abstract and difficult to imagine."
"We have to be aware of the dangers of quantum technology."

\*\*\**Questions to deepen the discussion:*
Did the        … experiments have an effect?
                      … lunch talk have an effect?
                      … presentations have an effect?
                      … other?

## Appendix C: Gender differences

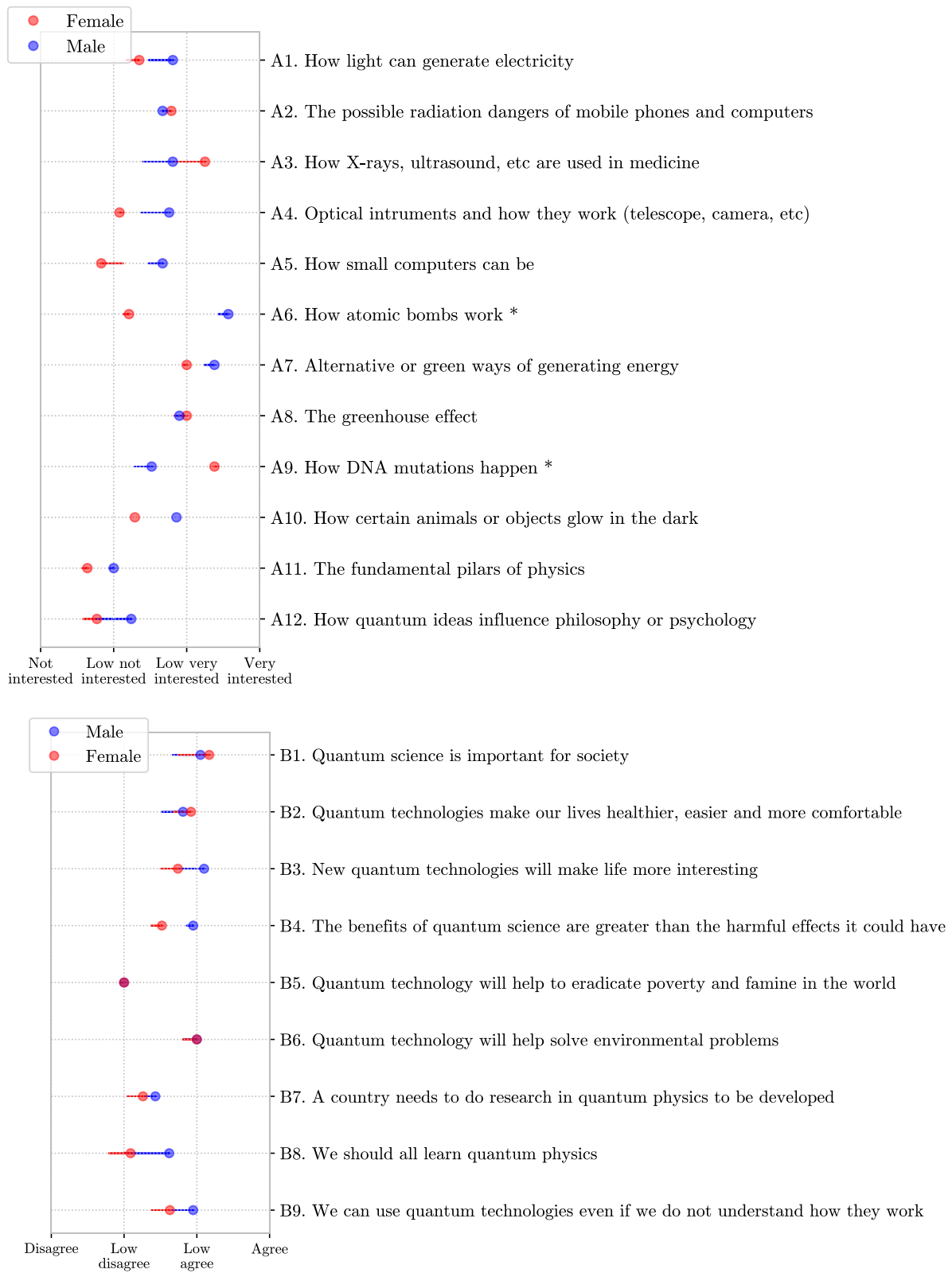

**Figure A1.** Averages for interests and opinions of girls and boys. The circles indicate the averages in the post-tests and the tails start at the averages of the pre-test. Gender differences were significant for statements A6 (p-value less than 0.001 in both tests), A9 (p-value less than 0.001 in pre-test and 0.002 in post- test), B4 (p-value 0.009 in pre-test), A4 (p-value 0.003 in post-test) and A5 (p-value 0.005 in post-test). Only A6 and A9 were significantly different in both tests.